\documentclass[a4paper]{jpconf}
\usepackage{graphicx}
\newcommand{\amev}{\textit{A} MeV\ }
\begin{document}
\title{Correlations between isospin dynamics and Intermediate Mass Fragments emission 
time scales:\\ a probe for the symmetry energy in asymmetric nuclear matter}

\author{E.~De~Filippo$^1$, F.~Amorini$^2$, L.~Auditore$^3$, V. Baran$^4$, I.~Berceanu$^5$, G.~Cardella$^1$,
M.~Colonna$^2$, E.~Geraci$^{6,1}$, S.~Gian\`i$^2$, L.~Grassi$^2$, A.~Grzeszczuk$^7$, P.~Guazzoni$^8$, J.~Han$^2$, 
E.~La~Guidara$^1$, G.~Lanzalone$^{9,2}$, I.~Lombardo$^{10}$, C.~Maiolino$^2$, T.~Minniti$^3$, 
A.~Pagano$^1$, M.~Papa$^1$, E.~Piasecki$^{11,12}$, S.~Pirrone$^1$, G.~Politi$^{6,1}$, A.~Pop$^5$, 
F.~Porto$^{6,2}$, F.~Rizzo$^{6,2}$, P.~Russotto$^1$, S.~Santoro$^3$, A.~Trifir\`o$^3$, M.~Trimarchi$^3$,
G.~Verde$^1$, M.~Vigilante$^{10}$, J.~Wilczy\'nski$^{12}$ and L.~Zetta$^8$}
\address
{
$^1$ INFN, Sezione di Catania, Italy \\
$^2$ INFN, Laboratori Nazionali del Sud, Catania, Italy, \\
$^3$ INFN, Gruppo Collegato di Messina and Dip. di Fisica, Univ. di Messina, Italy\\
$^4$ Physics Faculty, University of Bucharest, Romania\\
$^5$ National Institute of Physics and Nuclear Engineering "Horia Hulubei", Bucharest, Romania\\
$^6$ Dipartimento di Fisica e Astronomia, Univ. di Catania, Catania, Italy\\
$^7$ Institute of Physics, University of Silesia, Katowice, Poland\\
$^8$ INFN, Sezione di Milano and Dipartimento di Fisica, Univ. di Milano, Italy\\
$^9$ ``Kore'' Universit\`a, Enna, Italy\\
$^{10}$ INFN, Sezione di Napoli and Dipartimento di Fisica, Univ. di Napoli, Italy\\
$^{11}$ Heavy Ion Laboratory, University of Warsaw, Warsaw, Poland \\
$^{12}$ National Centre for Nuclear Research, Otwock-\'Swierk, Poland
}
 
\ead{defilippo@ct.infn.it}

\begin{abstract}
We show new data from the $^{64}$Ni+$^{124}$Sn and $^{58}$Ni+$^{112}$Sn reactions 
studied in direct kinematics with the CHIMERA detector at INFN-LNS 
and compared with the reverse kinematics reactions at the same incident 
beam energy (35 \amev). Analyzing the data with the method of relative velocity 
correlations, fragments coming from statistical decay of an excited projectile-like 
(PLF) or target-like (TLF) fragments are discriminated from the ones coming 
from dynamical emission in the early stages of the reaction.  
By comparing data of the reverse kinematics experiment
with a stochastic mean field (SMF) + GEMINI calculations our results show
that observables from neck fragmentation mechanism add valuable
constraints on the density dependence of symmetry energy. 
An indication is found for a moderately stiff symmetry energy potential 
term of EOS.
\end{abstract}

\section{Introduction}
Heavy ion collisions are a powerful tool to study the nuclear properties at different 
conditions of density, temperature and isospin asymmetry. For example, in the Fermi energy regime 
(10-100 \amev), it is possible to explore from very low barionic densities ($\rho / \rho_0\le0.2$) 
clustered matter \cite{wad12} to densities in proximity to the saturation value 
($\rho_0 = 0.16 fm^{-3}$) \cite{bar04}. Conversely, in the relativistic energy 
regime (E/A $\ge$ 100 \amev) it is potentially possible to access densities up to $2-2.5$ 
times the saturation value in a short timescale \cite{tra12,rus11}. For these reasons it is not 
surprising that heavy ion collisions, in particular with projectiles and targets with large 
isospin asymmetries, have been widely used to probe the density dependence of the 
symmetry term of the nuclear Equation of State (EOS), that is a key ingredient for the  
dynamical model calculations of heavy ion collisions and astrophysical predictions \cite{hor01}. 
Different experimental observables have been used to constraint the density dependence of the symmetry 
energy in heavy-ion collisions: isospin diffusion and equilibration \cite{tsa09}, 
neutron to proton ratio \cite{fam06}, light charged particles transverse collective flow 
\cite{koh11}, ratio of fragments yields and isoscaling \cite{mar12}, heavy residues production 
in semi-central collisions \cite{amo09}, isospin migration 
in the low density ``neck'' region \cite{def12}. These different observables from heavy 
ion reactions at low densities are generally consistents within uncertainties \cite{tsa12}, 
but results can be strongly dependent from comparison with microscopic calculations simulating 
the nuclear dynamics. Indeed a weak overlap exists so far between constraints from heavy 
ion collisions and the estimate of the symmetry energy from astrophysical observations 
on neutron stars \cite{ste12,tsa12}. 
For these reasons it is important to add new observables, increasing accuracies of the existing ones and to 
improve the model simulations in order to reduce the relative large actual uncertainties.        

In this contribution we present experimental data for the $^{64}$Ni+$^{124}$Sn and $^{58}$Ni+$^{112}$Sn 
reactions studied in direct kinematics with the CHIMERA detector at the same beam incident 
energy (35 \amev) of the previously studied experiment in reverse kinematics \cite{def05,def12}.  
The ensemble of data of the two experiments collects a unique set of information on the midrapidity
``neck'' fragmentation mechanism in semi-peripheral dissipative collisions. 
We show that the Intermediate Mass Fragments (IMF, 3$\le$Z$\le$20) midrapidity emission 
presents many experimental properties (like the N/Z isospin asymmetry enhancement) 
that in transport models calculations are attributed to the 
formation, in the early stage of the reaction, of a low density region (``neck'') 
connecting projectile-like and target-like fragments. 
These properties can be linked to reaction dynamics if they 
are correlated with the timescale evolution of the nuclear reactions. In this contest we have 
compared experimental data for the reverse kinematics experiment with a stochastic mean field 
calculation (SMF) in order to get a parametrization for the potential symmetry energy 
term of EOS. 

\section{Experimental details}
The direct kinematic experiment has been performed at the Catania INFN-LNS superconducting cyclotron 
where beams of $^{64}$Ni and $^{58}$Ni at 35 A.MeV of ≈0.5 enA intensity impinged on self-supporting 
$^{124}$Sn and $^{112}$Sn thin targets whose thickness are respectively 187 $\mu g/cm^2$ and 268  $\mu g/cm^2$. Reaction products were detected with the 4$\pi$ detector CHIMERA in its full configuration, 
constituted by 1192 Si-CsI(Tl) telescopes arranged in 35 rings around the beam axis. 
Data acquisition was triggered when at least three silicon detectors were fired. 
This experiment was the first to use on a large number of detectors the CHIMERA silicon pulse shape 
upgrade \cite{ald05}, giving the possibility to charge identify the particles that are stopped in 
silicon detectors. In the reverse kinematics experiment only the forward part of the 
detector (688 telescopes covering the angular range between 1$^o$ and 30$^o$) was used. Details of this 
latter experiment setup are described elsewhere \cite{def05,def12}.

\begin{figure}
\begin{centering}
\includegraphics[width=0.55\textwidth]{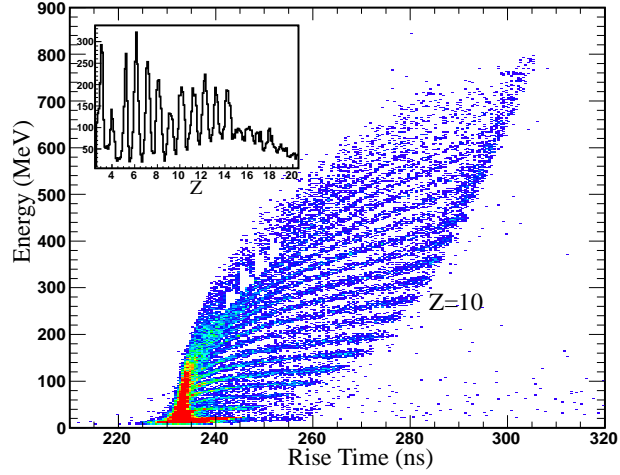}
\caption{Rise-time of impinging particles vs. particles energy bi-dimensional plot
for particle stopping in a silicon detector at 13.75$^o$ for the reaction $^{64}$Ni+
$^{124}$Sn. The inset shows the quality of charge identification obtained.}
\label{fig1}
\end{centering}
\end{figure}

Fig. \ref{fig1} shows an example of the charge identification obtained for light 
fragments (Z$\le$20) using the pulse shape discrimination (PSD) technique for a silicon detector. 
The identification threshold is around 4 A.MeV for light charges (Z=6). The use of PSD technique 
was very important in this direct kinematics experiment because a part of slow IMFs, produced 
in the mid-rapidity region, fall below the threshold for $\Delta E-E$ identification. 
We have selected almost complete events where the total charge is 45$\le Z_{TOT}\le$80 
and the parallel momentum of the colliding system is at least 60\% of the total one. 
Semipheripheral collisions were selected gating on the total charged particle multiplicity M$\le$7. 
Following the same methods used in Refs. \cite{def05,def12} in order to select ternary events 
in the final state, we considered a subset of events where the total charge of the three biggest 
fragment Z(1)+Z(2)+Z(3) $\ge$ 45 and their momentum $p(1)+p(2)+p(3)\ge$ 0.6$p_{beam}$.  

\begin{figure}
\begin{centering}
\includegraphics[width=0.5\textwidth]{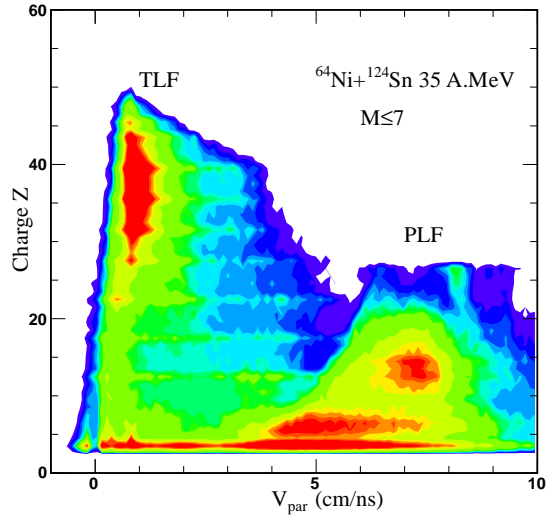}
\caption{Atomic number of fragments Z as a function of their parallel velocity along the 
beam axis $V_{par}$ for the selected ternary events. PLF and TLF regions are indicated 
by labels.}
\label{fig2}
\end{centering}
\end{figure}

Fig. \ref{fig2} shows for the selected events the 2-dimensional plot of fragments parallel velocity 
along the beam axis as a function of their charge for the $^{64}$Ni+$^{124}$Sn reaction.
It is interesting to compare this plot with the same obtained for the reverse kinematics 
experiment (Fig. 2 of Ref. \cite{def05}). Here the slow-moving and massive TLF fragments are 
produced around 1 $cm/ns$ velocity and all particles belonging to this region (mainly stopping in 
the first stage silicon detectors) are identified in mass with Time-of-Flight techniques. 
The complete angular coverage at least up to 90$^o$ becomes essential to study events in 
which projectile-like and target-like fragments are simultaneously present in the same events 
for semi-peripheral collisions. We clearly see also the two regions respectively of PLF (Ni-like) 
with velocities around 7.5 cm/ns and of IMFs in the intermediate velocity region between 
the two main partners of the reaction.  

\section{Results}
\begin{figure}[b]
\begin{centering}
\includegraphics[width=0.55\textwidth]{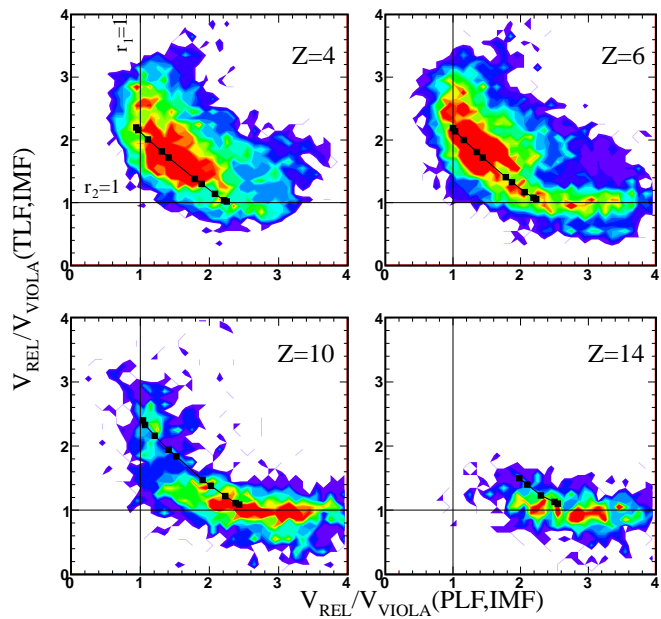}
\caption{For the $^{64}$Ni+$^{124}$Sn reaction, correlations between relative velocities 
$V_{REL}/V_{Viola}$ of the three biggest fragments in the event for $Z_{IMF}$=4,6,10,14. 
The points refers to a monodimensional Coulomb trajectory calculation (see text).}
\label{fig3}
\end{centering}
\end{figure}

In order to disentangle dynamically and statistically emitted fragments and to look for timescale 
of fragment formation, the three biggest fragments of each event were sorted according to the 
decreasing value of their parallel velocity, following the method described in \cite{def12}. 
The three particles, labeled as PLF,IMF (3$\le$Z$\le$20) or TLF, 
depending upon their respective velocity were analyzed to check the correct attribution event by event; 
finally the fragment-fragment relative velocities $V_{REL}(PLF,IMF)$ and $V_{REL}(TLF,IMF)$  
were calculated and are reported in Fig. \ref{fig3}.    

The relative velocities are normalized to the one corresponding to the Coulomb repulsion as given 
by the Viola systematics \cite{hin87}. Fig. \ref{fig3} shows the correlations between 
the two relative velocities  $r_1$=$V_{REL}(PLF,IMF)/V_{Viola}(PLF,IMF)$ and 
$r_2$=$V_{REL}(TLF,IMF)/V_{Viola}(TLF,IMF)$ for the IMFs charges Z=4,6,10,14.  
As compared with a similar plot in the Reverse experiment 
(Ref. \cite{def05}, fig. 7) we note now that
we can populate with similar efficiency both the regions 
along the $r_1$=1 axis (whose yield is dominated by sequential emission from PLF) and $r_2$=1 axis  
(whose yield is mainly due to sequential emission from TLF); values of $r_1$ and $r_2$ 
simultaneously larger than unity indicate a prompt ternary division (dynamical origin). 
We can observe that heavier fragments (as Z=14 in the figure) are originated mainly by the break-up 
or fission of the heavy partner, i.e. the target-like residue in our case and 
they lie along the r$_2$=1 axis, 
while light fragments are concentrated along the diagonal, indicating their prevailing dynamic origin. 
In Fig. \ref{fig3} a timescale calibration was done, 
as in Ref. \cite{def05} using a three-body collinear Coulomb trajectory calculation. 
The inner points along the diagonal correspond to the shortest timescales (40-60 fm/c), 
corresponding to IMFs predominantly emitted from the dynamically expanding neck region formed 
at midrapidity, between the projectile-like and target-like primary fragments. This method permits to explore 
and disentangle different stages of the dynamical evolution of the system, that normally are mixed 
together when looking, for example in a source recognition analysis, at Galilean invariant 
$V_\parallel-V_\perp$ contour diagrams.     

In order to study the alignment properties of midvelocity fragments, we have evaluated 
the angle $\theta_{PROX}$. As shown in the inset of Fig. 4d), if the IMF had its origin from a 
PLF break-up, $\theta_{PROX}$ is the angle between the (PLF-IMF center of mass)-TLF relative 
velocity axis, and the PLF-IMF break-up axis (relative velocity between PLF and IMF oriented from the light 
to the heavy fragment). This definition mainly differs from the ones presented 
in Refs. \cite{boc00,hud12} because it requires the explicit detection of a TLF and PLF fragments
in the same event.  
$\cos(\theta_{PROX})$=1 indicates a complete alignment with 
the IMF emitted in the backward hemisphere respect to the PLF and $\cos(\theta_{PROX})<0$ 
indicates the emission in the forward hemisphere respect to the PLF. Of course 
the IMF emission could be attributed also to a TLF break-up. As a first approximation, 
we have calculated $\cos(\theta_{PROX})$ requiring the condition
$V_{REL}(PLF,IMF)/V_{Viola}<$1.6. This condition removes most of IMFs 
emitted by TLF \cite{rus12}. A more complete analysis in the direct kinematics experiment 
will give the possibility to extend and complete these results considering 
the contribution due to both the TLF and PLF break-up. 
 
\begin{figure}
\begin{centering}
\includegraphics[width=0.62\textwidth]{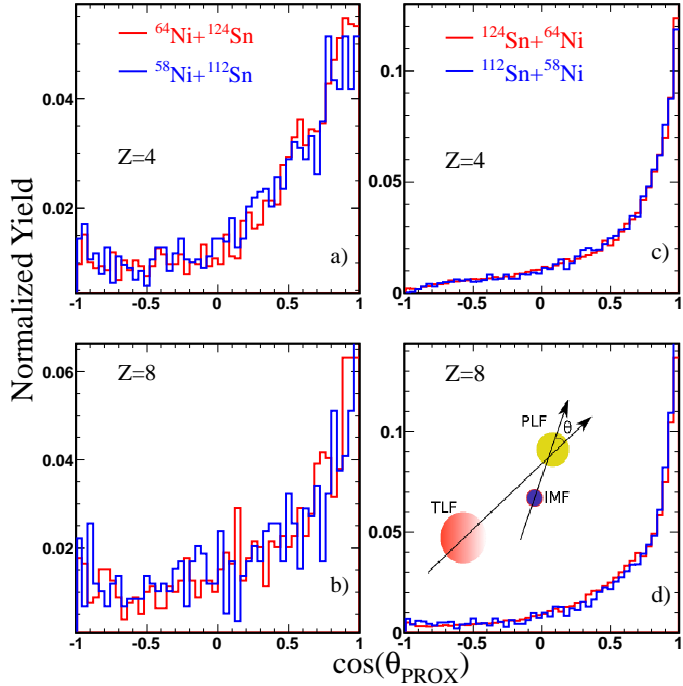}
\caption{a),b) panels: $\cos(\theta_{PROX})$ angular distribution calculated for a PLF 
(Ni-like) break-up and production of Z=4 and Z=8 IMFs for the reactions $^{64}Ni+^{124}Sn$ 
(red histogram) and $^{58}$Ni+$^{112}$Sn (blue histogram); c), d) panels: same 
distributions for the for the reverse kinematic reactions $^{124}$Sn+$^{64}$Ni (red histogram) and  
$^{112}$Sn+$^{58}$Ni (blue histogram) for a PLF (Sn-like) break-up. The insert in Fig. 4d) 
gives a sketch of the $\theta_{PROX}$ definition. }
\label{fig4}
\end{centering}
\end{figure}

Fig \ref{fig4} shows the $\cos(\theta_{PROX})$ angular distribution for Z=4 and Z=8 IMFs for the 
two reactions under study in the direct kinematic reaction (a,b left panels) and in the reverse
kinematic reaction (c,d right panels) respectively. The yields are normalized to the respective 
area. For a statistical emission the $\cos(\theta_{PROX})$ distribution is expected to show 
a forward-backward symmetry around $\cos(\theta_{PROX})=0$. We note that the distributions 
are peaked at $\cos(\theta_{PROX})\approx 1$ indicating a strong 
anisotropy favoring the backward emission respect to the forward one in a strict aligned 
configuration along the TLF-PLF separation axis.    
In Ref. \cite{def12} has been shown in details how, adding a kinematic condition on the fragments degree 
of alignment to a condition that selects the greatest deviations from Viola systematics in 
the relative velocities correlation plots (as illustrated in Fig. \ref{fig3}), it is possible 
to careful disentangle the pattern of dynamically emitted fragments (characterized by short emission times 
and by a strong degree of alignment in $\cos(\theta_{PROX})$), respect to statistically emitted fragments. 
For these two classes of events it is now possible to introduce a further observable, i.e. the fragments 
isotopic composition N/Z. This is the object of the last section.         

\begin{figure}[b]
\begin{centering}
\includegraphics[width=0.67\textwidth]{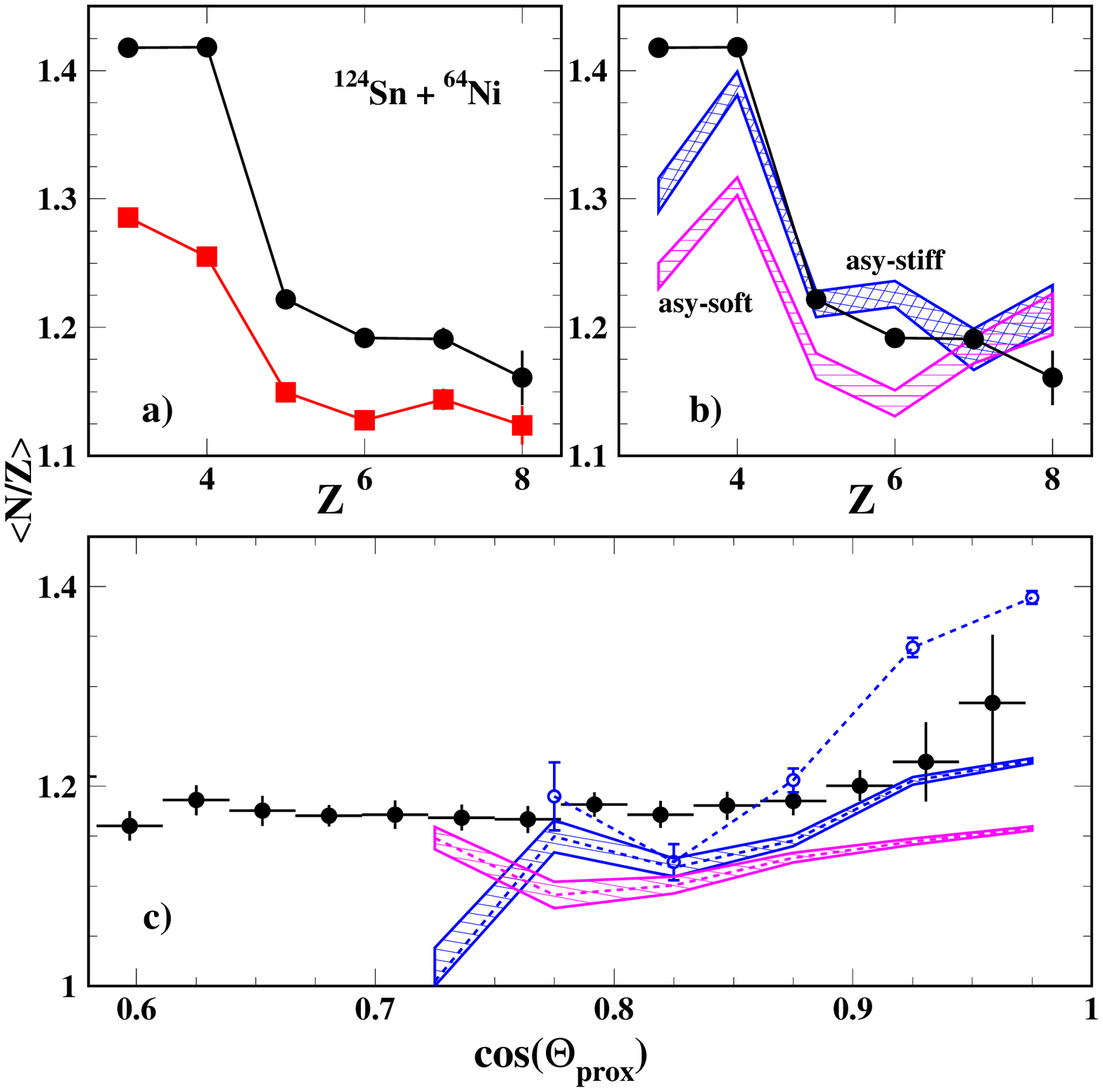}
\caption{a) For the $^{124}$Sn+$^{64}$Ni reaction, experimental $<N/Z>$ distribution of IMFs as a function
of charge Z for dynamically emitted particles (solid circles) and statistically emitted particles 
(solid squares); b) solid circles: same experimental data of Fig. 5a) for dynamically emitted particles. 
Blue hatched area: SMF-GEMINI calculations for dynamically emitted particles and 
asy-stiff parametrization; magenta hatched area: asy-soft parametrization for dynamically emitted particles. 
c) solid circles: experimental $<N/Z>$ as a function of $\cos(\theta_{PROX})$ for charges $5\le Z \le 8$; empty circles: SMF calculation for primary fragment (asy-stiff parametrization); SMF-GEMINI calculations 
are indicated by blue-hatched area (asy-stiff parametrization) and magenta hatched area 
(asy-soft) respectively.}
\label{fig5}
\end{centering}
\end{figure}

\section{Isotopic distributions and comparisons with SMF model}
Fig. \ref{fig5}a) shows the $<N/Z>$ as a function of the IMFs atomic number Z for the 
reaction $^{124}$Sn+$^{64}$Ni. Fragments statistically emitted in the PLF forward hemisphere have been 
selected by using the condition $\cos(\theta_{PROX})<0$. The relative points are shown as solid squares 
in Fig. \ref{fig5}a). Solid circles shows the $<N/Z>$ for dynamically emitted IMFs. Events for these particles 
are selected by imposing that $\cos(\theta_{PROX})>0.8$ (highest enhancement for backward emission) and 
selecting events near the diagonal in the $V_{REL}/V_{Viola}(PLF,IMF) - V_{REL}/V_{Viola}(TLF,IMF)$ 
relative velocities correlation plot. We note that the $<N/Z>$  ratio is systematically 
larger for dynamically emitted particles respect to the statistically emitted ones. In Fig. \ref{fig5}c)
we have reported (solid circles), for the same reaction, 
the correlation between $\cos(\theta_{PROX})$ and $<N/Z>$ for all 
fragments with charges between $5\le Z \le 8$. We observe an increase of the $<N/Z>$ at values of 
$\cos(\theta_{PROX})$ approaching to $1$, corresponding to the highest degree of alignment. A similar 
result has been found recently, in a different data analysis contest, 
for the $^{124}$Xe+$^{124,112}$Sn system \cite{hud12}. 

The data, for the inverse kinematics neutron rich reaction $^{124}$Sn+$^{64}$Ni, 
were compared with a transport theory using the stochastic mean field model (SMF) 
based on Boltzmann-Norheim-Vlasov (BNV) equation \cite{dit10,col98}. The potential part 
of the symmetry energy is taken into account using two different parametrizations as a function of 
density named \textit{asy-stiff} and \textit{asy-soft}. The first one linearly increases 
with the density while the second one exhibits a weak variation around the 
saturation density $\rho_0$. The slope parameter of the symmetry energy, defined as
\begin{equation}
L=3\rho_0 \left. {dE_{sym}(\rho) \over d\rho}\right|_{\rho=\rho_0}
\end{equation}       
is in the current calculation around 80 MeV for the asy-stiff and 25 MeV for the asy-soft 
choice. The statistical code GEMINI \cite{char88} is used as a second step de-excitation phase 
applied to the SMF primary hot fragments. 

\begin{figure}
\begin{centering}
\includegraphics[width=0.50\textwidth]{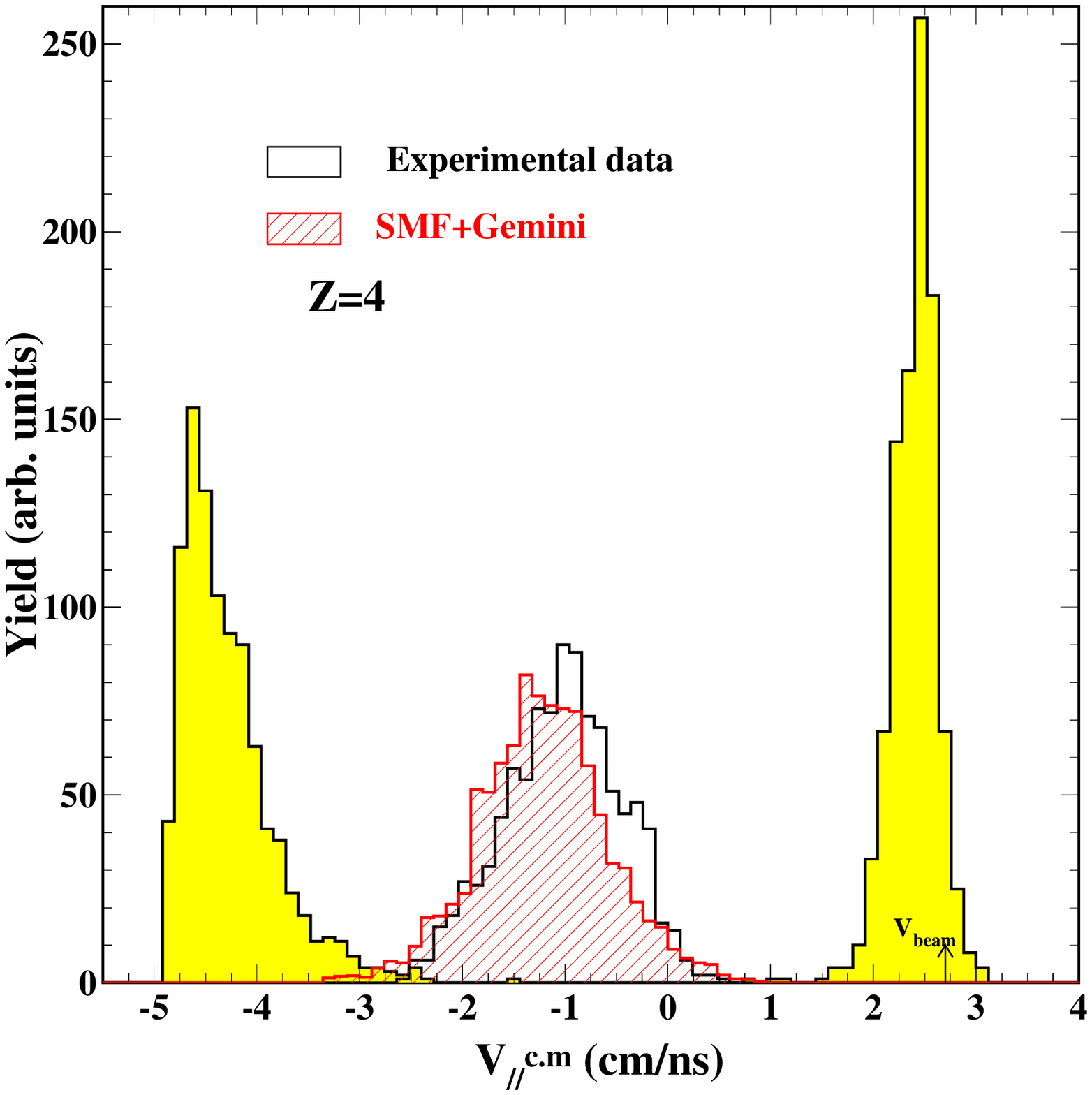}
\caption{For the $^{124}$Sn+$^{64}$Ni reaction, black empty histogram: velocity spectra $V_\parallel$ in the 
c.m. system of dynamically emitted fragments with charge Z=4. Hashed red histogram: calculated 
velocity spectra (SMF+GEMINI) for the same reaction. Calculated data are filtered for detector 
acceptance. Experimental velocity distributions (yellow histograms) of PLF and TLF residues,  
detected in the same ternary event are plotted as reference. }
\label{fig6}
\end{centering}
\end{figure}

In Figs. \ref{fig5}b,c) the SMF + GEMINI calculations are plotted as hatched area histograms for 
dynamically emitted fragments. The hatched zone indicates the error bars in the calculation.
As can be observed in Fig. \ref{fig5}b) the asy-stiff 
parametrization (blue hatched area) produces more neutron rich fragments respect to the asy-soft 
choice and the difference persists after the GEMINI secondary-decay stage for Z$<$7. 
The asy-stiff parametrization matches the experimental data fairly well. This is confirmed in Fig.  
\ref{fig5}c) where the asy-stiff parametrization better reproduces the $<N/Z>$ enhancement observed 
for values of $\cos(\theta_{PROX})>0.9$. In the calculations the dynamical emitted fragments 
production yields become near to zero below $\cos(\theta_{PROX})<$0.7, thus theoretical $<N/Z>$   
values can not be calculated for $\cos(\theta_{PROX})$ below $\approx$0.7.  
In Fig. \ref{fig5}c) SMF calculations for primary fragments 
and asy-stiff parametrization are also shown (empty circles symbol). 

We have checked if the SMF + GEMINI calculation with asy-stiff parametrization 
reproduces other basic experimental observables. As an example, we show in Fig. \ref{fig6} 
the experimental longitudinal velocity distribution respect to beam axis 
(in the c.m. system) of Z=4 dynamically emitted fragments (empty histogram). 
Velocity distributions (yellow histograms) of experimental PLF and TLF fragments,  
detected in the same ternary event are drawn as reference in the same plot.
Data are compared with results of SMF+GEMINI calculation for Z=4 neck emitted fragments 
(hashed red histogram). Calculated data are corrected for the detector geometrical acceptance,
thresholds and time resolution. The shape and position of the experimental data are well 
reproduced for all IMFs with charges from Z=3 to Z=8 (only charge Z=4 is shown in the plot). 
More details on calculations and comparison with experimental data can be found in \cite{def12}.        

\section{Summary and outlook}
We have studied with the 4$\pi$ detector CHIMERA the two reactions $^{64,58}$Ni+$^{124,112}$Sn  
and  $^{124,112}$Sn + $^{64,58}$Ni  at the same energy of relative motion (35 \amev).
We defined a method to disentangle sequentially from dynamically emitted particles at midrapidity and 
to correlate the isotopic composition of intermediate mass fragments with their emission timescale.  
Dynamically emitted IMF shows larger values of $<$N/Z$>$ isospin asymmetry and stronger angular anisotropies  
supporting the concept of ``isospin migration'' in neck fragmentation mechanism. 
Comparing the data for the reverse kinematics reaction with a stochastic mean field simulation, 
valuable constraints on the symmetry energy term of nuclear EOS at sub-saturation densities is obtained. 
A stiff $E_{sym}(\rho)$ behavior, with $L\approx$80 MeV corresponding to a linear density dependence of the 
symmetry energy, better reproduces our data. All these aspects open new perspectives for reaction studies with 
exotic beams. Our first outlook is in fact the possibility to plan new experiments using a 30 \amev $^{68}Ni$ 
beam recently produced at LNS \cite{pag12}. As a second outlook, we have recently proposed to study the  
$^{124}$Xe+$^{64}$Zn reaction as compared with $^{124}$Sn+$^{64}$Ni system where only 
the N/Z changes for the two systems with the same masses. We hope this study will permit to disentangle 
mass from isospin asymmetry effects evidencing the effective role of symmetry energy 
in the dynamics of the reactions.


\section*{References}
\begin{thebibliography}{99}
\bibitem{wad12} Wada R et al. (2012) \textit{Phys. Rev. C} \textbf{85} 064618 
\bibitem{bar04} Baran V et al. (2004) \textit{Nucl. Phys. A} \textbf{730} 329 
\bibitem{tra12} Trautmann W and Wolter H H (2012) \textit{Int. J. Mod. Phys. E} \textbf{21} 1230003 
\bibitem{rus11} Russotto P et al. (2011) \textit{Phys. Lett. B} \textbf{697} 471 
\bibitem{hor01} Horowitz C J and Piekarewicz J (2001) \textit{Phys. Rev. C} \textbf{64} 062802 
\bibitem{tsa09} Tsang M B et al. (2009) \textit{Phys. Rev. Lett.} \textbf{102} 122701 
\bibitem{fam06} Famiano M A et al. (2006) \textit{Phys. Rev. Lett.} \textbf{97} 052701 
\bibitem{koh11} Kohley Z et al. (2011) \textit{Phys. Rev. C} \textbf{83} 044601
\bibitem{mar12} Marini P et al. (2012) \textit{Phys. Rev. C} \textbf{85} 034617 
\bibitem{amo09} Amorini F et al. (2009) \textit{Phys. Rev. Lett.} \textbf{102} 112701 
\bibitem{def12} De Filippo E et al. (2012) \textit{Phys. Rev. C} \textbf{86} 014610 
\bibitem{tsa12} Tsang M B et al. (2012) \textit{Phys. Rev. C} \textbf{86} 015893 
\bibitem{ste12} Steiner A W and Gandolfi S (2012) \textit{Phys. Rev. Lett.} \textbf{108} 081102 
\bibitem{def05} De Filippo E et al. (2005) \textit{Phys. Rev. C} \textbf{71} 044602 
\bibitem{ald05} Alderighi M et al.  (2005) \textit{IEEE Trans. Nucl. Science} \textbf{52} 1624
\bibitem{hin87} Hinde D J et al. (1987) \textit{Nucl. Phys. A} \textbf{472} 318 
\bibitem{rus12} Russotto P et al. (2010) \textit{Phys. Rev. C} \textbf{81} 064605 and 
Russotto P et al. \textit{(2012), to be submitted} 
\bibitem{boc00} Bocage F et al. (2000) \textit{Nucl. Phys. A} \textbf{676} 391 
\bibitem{hud12} Hudan S et al. (2012) \textit{Phys. Rev. C} \textbf{86} 021603(R) 
\bibitem{dit10} Di Toro M et al. (2010) \textit{J. Phys. G} \textbf{37} 083101 
\bibitem{col98} Chomaz P, Colonna M and Randrup J (2004) \textit{Phys. Rep.} \textbf{389} 263 
\bibitem{char88} Charity R et al. (1988) \textit{Nucl. Phys. A} \textbf{483} 371 
\bibitem{pag12} Pagano A (2012) \textit{Proceedings of the IWM2011, Int. Workshop on Multifragmentation 
and Related Topics}, Caen (France) 2-5 November 2011, EPJ Web of Conferences vol. 31, 00005 
\end{thebibliography}
\end{document}